%% file: quarks.tex
\documentclass[12pt]{article}
\usepackage{amsmath,amssymb,amsthm}
\usepackage{bbm,latexsym}
\usepackage{graphicx}
\usepackage{rotating} 
\usepackage[numbers,sort&compress]{natbib}
\newcommand{\diag}{\mbox{diag}}
\begin{document}
\title{\normalsize \hfill UWThPh-2015-2 \\[1cm] \LARGE
A complete survey of texture zeros \\ 
in general and symmetric quark mass matrices}
\author{P.O. Ludl\thanks{E-mail: patrick.ludl@univie.ac.at} 
\setcounter{footnote}{6}
and W. Grimus\thanks{E-mail: walter.grimus@univie.ac.at} \\[4mm]
\small University of Vienna, Faculty of Physics \\
\small Boltzmanngasse 5, A--1090 Vienna, Austria \\[4.6mm]}

\date{January 20, 2015}

\maketitle

\begin{abstract}
We perform a systematic analysis of all possible texture zeros in
general and symmetric quark mass matrices. Using the values of 
masses and mixing parameters at the
electroweak scale, we identify for both cases
the maximally restrictive viable textures. Furthermore, we investigate the
predictive power of these textures by applying a
numerical predictivity measure recently defined by us. With this
measure we find no predictive textures
among the viable general quark mass
matrices, while in the case of symmetric quark mass matrices 
most of the 15 maximally restrictive textures 
are predictive with respect to one or more light quark masses.
\end{abstract}

\newpage

One of the most interesting questions of flavour physics is whether
mixing angles are related to fermion mass ratios, like in the famous
relation~\cite{gatto}
\begin{equation}\label{c}
\sin \theta_c \simeq \sqrt{\frac{m_d}{m_s}}
\end{equation}
between the Cabibbo angle $\theta_c$ and the ratio of down-quark mass
to strange-quark mass. Here and in the following,  
quark masses are denoted by 
$m_q$ with $q = d,s,b$ for the
down-type quark masses
and $q =u,c,t$ for the
up-type quark masses.
It is an open question if equation~(\ref{c}) is only an empirical
relation or if there is a deeper reason for it founded in a hitherto
undiscovered theory of flavour.
Obviously, since the CKM matrix $U$ is defined as
\begin{equation}
U = {U^{(u)}_L}^\dagger U^{(d)}_L
\end{equation}
with diagonalization matrices $U^{(u)}_L$ and $U^{(d)}_L$
given by 
\begin{equation}
{U^{(d)}_L}^\dagger M_d U^{(d)}_R = \diag\left( m_d, m_s, m_b \right)
\quad\mbox{and}\quad
{U^{(u)}_L}^\dagger M_u U^{(u)}_R = \diag\left( m_u, m_c, m_t \right)
\end{equation}
for the down-type and up-type quarks, respectively,
the quark mass matrices $M_d$
and $M_u$ must have some structure in order to deduce relations like
equation~(\ref{c}). 
The simplest attempt to achieve such relations is to place texture
zeros in
the mass matrices~\cite{fritzsch}. 
Apart from simplicity, texture zeros have the feature that they are
practically synonymous with Abelian symmetries~\cite{grimus}.
Unfortunately, even in
this limited framework
no clear-cut predictive model has emerged---see for
instance~\cite{reviews,serodio} 
for reviews
and~\cite{sharma} for an attempt on
a unified texture in both quark and lepton sector. Therefore, it is 
appropriate to perform a complete study of all possibilities, as was
recently done in~\cite{ludl2014} for the lepton sector (see
also~\cite{lavoura}). 

In the analysis of~\cite{ludl2014} the notion of 
``maximally restrictive'' textures plays an important role. These have
a maximal number of zeros in the pair $(M_d,M_u)$ in the sense 
that by placing one more zero into this pair it becomes
incompatible with experimental data. 
It turned out that in the lepton sector the
predictive power
of general mass matrices with texture zeros is rather limited even for
maximally restrictive textures. Actually, 
we find
the same in
the quark sector, as we will explain in more detail below. In view of
this result, 
we perform 
an additional
analysis with 
\emph{symmetric} quark mass matrices.\footnote{In many papers 
  the mass matrices are assumed to be Hermitian. It is true that
  one can always achieve this by separate weak-basis transformations
  on the right-handed
  quark fields,
  however, this is not
  a valid argument because in the first place it is the texture zeros
  which define a basis and performing a subsequent basis transformation will in
  general remove the texture zeros.\label{footnote-hermitian}} 
In this paper, this is not more
than a facile assumption in order to enhance
predictivity,\footnote{For this purpose, 
other assumptions are possible as well, like for instance a
scaling ansatz~\cite{ghosal}.} 
however, it
could be motivated by left-right symmetric models~\cite{pati} or by
models based on $SO(10)$---see~\cite{chen} for reviews---with
renormalizable Yukawa couplings to scalar 10-plets and 126-plets, but
not to scalar 120-plets which would introduce an antisymmetric
component in the mass matrices~\cite{kuehboeck}.

Under weak-basis transformations the mass matrices transform as
\begin{equation}\label{wb}
M_d \to V_L^\dagger M_d V_R^{(d)}, \quad
M_u \to V_L^\dagger M_u V_R^{(u)}
\end{equation}
with unitary matrices $V_L$, $V_R^{(d)}$, $V_R^{(u)}$. Such
transformations have no effect on the quark masses and the mixing
matrix, \textit{i.e.}\ they do not change the physical predictions. 
These weak-basis transformations are used in two ways.
Firstly, it is well-known
that with weak-basis transformations of equation~(\ref{wb}) 
one can generate some zeros in 
$(M_d,M_u)$ which have, therefore, no predictive power at
all~\cite{weak-basis}. Such cases we exclude a priori from our
analysis---for more details on
this issue see~\cite{ludl2014}. 
Secondly, if one wants to preserve the 
zeros in the mass matrices, the unitary matrices occurring in
equation~(\ref{wb}) have to be restricted to permutation
matrices times diagonal matrices of phase factors. 
Weak-basis transformations where the unitary matrices in
equation~(\ref{wb}) are pure permutation matrices,
\textit{i.e.}\ ``weak-basis permutations''~\cite{ludl2014}, 
allow to divide the possible patterns
of texture zeros in $(M_d,M_u)$ into equivalence classes with identical
predictions and it is thus sufficient to treat one representative 
$(M_d^{(i)},M_u^{(i)})$ of each equivalence class. 
Finally, with weak-basis transformations 
which are diagonal matrices of phase factors, \textit{i.e.}\ by
rephasing, one can then remove redundant phases from each representative 
$(M_d^{(i)},M_u^{(i)})$ in order to obtain representatives with the
minimal number of parameters.

Another important aspect of our general analysis is that we do not
consider model realizations of the textures. Therefore, we cannot
treat radiative corrections and we have to assume therefore that the
quark masses and the CKM matrix can be reproduced with sufficient
accuracy by tree-level mass matrices. Consequently, we take into account only
non-singular mass matrices. 

To test 
if a texture is compatible with 
the observations, we perform a $\chi^2$-analysis. 
We have ten physical observables: the six quark masses, the three
mixing angles and the CP-violating phase. We have to check for each  
texture $(M_d^{(i)},M_u^{(i)})$ whether it can
reproduce the input data within experimental errors.
Actually, for the mixing matrix $U$ we prefer to use the Wolfenstein
parameters~\cite{wolfenstein} $\lambda$, $A$, $\bar\rho$ and
$\bar\eta$, as defined in~\cite{rpp}. In order to have a consistent set
of input data, we have to fix a common energy scale $\mu$ at which 
the quark masses and mixing parameters are taken. 
We settle on the scale $\mu = M_Z$,
the mass of the $Z$ gauge boson, which means that, if the
texture zeros have a symmetry realization, then this symmetry is
effective at the
electroweak scale.\footnote{Within the Standard Model
the relevant mass ratios $m_d/m_b$, $m_s/m_b$, $m_u/m_t$
and $m_c/m_t$ do not significantly change due to renormalization
group running of the quark masses at scales from 
$\mu\sim 2\,\text{GeV}$
to $\mu \sim m_t$.
This can be easily checked using the
results of~\cite{xing-zhou}.
Moreover, within the Standard Model also the mixing angles do
not run significantly.
Therefore, if the effects of physics beyond
the Standard Model are small at scales $\mu \lesssim m_t$, the
analysis of texture zeros will not be affected by the exact choice of $\mu$.}
This scale has the advantage that
all observables used for the input, with the
exception of the top quark mass, are measured at energies below $M_Z$
and, therefore, are evolved by the renormalization group equations of
the Standard Model to the scale $\mu = M_Z$. 
Concretely, we take the input values at $M_Z$ 
from~\cite{antusch}; since in that paper the mixing angles and the CKM
phase are given,
we have to convert these into the Wolfenstein
parameters. We display our input in table~\ref{input}.
\renewcommand{\arraystretch}{1.4}
\begin{table}
\begin{center}
\begin{tabular}{|l|ll|}
\hline
$m_u / (10^{-6}v)$ & $7.4$ & $^{+1.5}_{-3.0}$ \\
$m_d / (10^{-5}v)$ & $1.58$ & $^{+0.23}_{-0.10}$ \\
$m_s / (10^{-4}v)$ & $3.12$ & $^{+0.17}_{-0.16}$ \\
$m_c / (10^{-3}v)$ & $3.60$ & $\pm 0.11$ \\
$m_b / (10^{-2}v)$ & $1.639$ & $\pm 0.015$ \\
$m_t / (10^{-1}v)$ & $9.861$ & $^{+0.086}_{-0.087}$ \\
\hline
$\lambda$ & $0.22540$ & $\pm 0.00070$ \\
$A$ & $0.828$ & $\pm 0.014$ \\
$\bar{\rho}$ & $0.133$ & $\pm 0.020$ \\
$\bar{\eta}$ & $0.350$ & $\pm 0.015$ \\
\hline
\end{tabular}
\caption{The quark masses and the parameters of the CKM matrix in the
  $\overline{\mathrm{MS}}$ scheme 
  at $\mu = M_Z$ computed within the Standard
  Model in~\cite{antusch}.
  The quark masses are given in units of $v=174.104\,\mathrm{GeV}$. 
  We have transformed the mixing parameters
  $\theta_{12},\theta_{23},\theta_{13}$ and $\delta$ 
  given in table~2 of~\cite{antusch} to the four Wolfenstein
  parameters~\cite{rpp} $\lambda$, $A$, $\bar{\rho}$ and $\bar{\eta}$
  assuming Gaussian error propagation. \label{input}}
\end{center}
\end{table}

Now we formulate a criterion that 
a texture $(M_d^{(i)},M_u^{(i)})$ is compatible with
the data. We stipulate that the contribution of each observable to
$\chi^2_\mathrm{min}$, the minimum of $\chi^2$, 
is at most $25$; this means that the deviation of the observable from its
experimental value is at most $5\sigma$~\cite{ludl2014}.
Since we have ten input values, 
this implies $\chi^2_\text{min} \leq 250$.

Even if we know that a texture $(M_d^{(i)},M_u^{(i)})$ is compatible with
the data, we do not know whether it has 
any predictive power or not. In order to discuss this question, we
apply the numerical method developed in~\cite{ludl2014} which we
repeat here briefly. 
The method is completely general, independent of
the problem under discussion.
Consider a model with a set of parameters $x$ making predictions
$P_j(x)$ for the observables $\mathcal{O}_j$ with 
experimental mean 
values $\overline{\mathcal{O}}_j$ and
errors $\sigma_j$.
Then the $\chi^2$-function of the model is given by\footnote{In case
  of asymmetric error 
intervals, $\sigma_j$ is replaced by $\sigma_j^{\mathrm{left}}$
and $\sigma_j^{\mathrm{right}}$ for $P_j(x) < \overline{\mathcal{O}}_j$
and $P_j(x) \geq \overline{\mathcal{O}}_j$, respectively.}
\begin{equation}\label{c2}
\chi^2(x) = \sum_j \chi^2_j(x)
\quad \mbox{with} \quad
\chi_j^2(x) = 
\left(
\frac{P_j(x)-\overline{\mathcal{O}}_j}{\sigma_j}
\right)^2.
\end{equation}
Let us assume that the model gives a good fit to the observables,
\textit{i.e.}\ $\chi^2_\mathrm{min}$ is sufficiently small.
Now we want to pose the question whether the model is predictive with
respect to the observable $\mathcal{O}_i$. Loosely speaking, this means
we want to investigate how much the prediction for $\mathcal{O}_i$ can
deviate from its mean value while varying $x$ such that 
all $P_j(x)$ with $j \neq i$ remain close to $\overline{\mathcal{O}}_j$.  
The numerical implementation is done in 
two steps~\cite{ludl2014}:
\begin{enumerate}
\item
We define 
\begin{equation}\label{ci}
\widetilde\chi_{i}^2(x) = \chi^2(x) - \chi_i^2(x),
\end{equation}
where we have removed the $\chi^2$-contribution of the observable whose
predictivity we want to investigate. With this 
$\widetilde\chi_{i}^2(x)$ we define a region in 
the parameter space via
\begin{equation}\label{def-B_i}
B_i =
\left\{ x\, \vert \,
\widetilde{\chi}_i^2(x) \leq \chi^2_\text{min} + \delta\chi^2
\quad \text{and} \quad
\chi_j^2(x) \leq 25 \enspace \forall j\neq i \right\},
\end{equation}
where $\chi^2_\mathrm{min}$ is the minimal value of the \emph{total}
$\chi^2(x)$ of equation~(\ref{c2}) and $\delta\chi^2$ is a
fixed parameter in the range $0 \leq \delta\chi^2 \lesssim 1$.
\item
With $B_i$ we formulate the predictivity measure for 
$\mathcal{O}_i$ as 
\begin{equation}\label{predictivity-measure}
\Delta(\mathcal{O}_i) = \max_{x \in B_i} \chi_i^2(x).
\end{equation}
\end{enumerate}
Note that for $x \in B_i$ the $P_j(x)$ with 
$j \neq i$ are within the $5\sigma$ region of their experimental mean values
$\overline{\mathcal{O}}_j$, 
\textit{i.e.}\ they are ``close'' to $\overline{\mathcal{O}}_j$
in the sense of our compatibility criterion
of a texture with the data.  
In equation~(\ref{def-B_i}), a non-zero $\delta\chi^2$ accelerates the
convergence of the numerical maximization of
$\Delta(\mathcal{O}_i)$~\cite{ludl2014}. In the present paper we have set 
$\delta\chi^2 = 0.1$. 

Clearly, the smaller $\Delta(\mathcal{O}_i)$ is, the better it is 
determined by the other observables. By choosing a bound
$b^2$, we can define a predictivity
criterion: for $\Delta(\mathcal{O}_i) \leq b^2$ we say the model is capable to
predict the observable $\mathcal{O}_i$; in this case, its value
deviates from its mean value by at most $b\sigma$. The choice of $b$
is rather arbitrary. We follow reference~\cite{ludl2014} and take
$b=10$. Thus our predictivity criterion is
\begin{equation}\label{bound}
\Delta(\mathcal{O}_i) \leq 100.
\end{equation}

The results of our analysis are presented in two tables,
table~\ref{general} for general mass matrices and
table~\ref{symmetric} for symmetric mass matrices.
In the general case, removing those pairs of mass matrices $(M_d,M_u)$
whose texture zeros can be generated by weak-basis transformations and
those with at least one singular matrix, we find 243 inequivalent
classes with representatives $(M_d^{(i)},M_u^{(i)})$.
Of these, 214 classes are compatible with the data and
among them there are 27
maximally restrictive classes whose representatives
$(M_d^{(i)},M_u^{(i)})$ are listed in table~\ref{general}. None of
these mass matrices are predictive in the sense discussed 
above.
\renewcommand{\arraystretch}{1.1}
\begin{table}
\begin{footnotesize}
\begin{center}
\input{classes_general.tex}
\end{center}
\end{footnotesize}
\caption{The 27 maximally restrictive textures in general quark mass matrices.
A matrix entry $0$ denotes a texture zero, and entries $1$ and $2$ stand for
real positive and complex parameters, respectively.
None of these textures is predictive with respect to any observable.
\label{general}}
\end{table}

For the symmetric mass matrices, 
we cannot apply general
weak-basis transformations which generate texture zeros, without
destroying the symmetry of the mass matrices.
Thus we discard only those pairs $(M_d,M_u)$ 
which have at least one singular matrix
and arrive in this way at 230 classes 
with representatives $(M_d^{(i)},M_u^{(i)})$. Out of these, 79 classes
survive the $\chi^2$-test. Finally, these classes contain 15 maximally
restrictive classes which are displayed in table~\ref{symmetric}.
According to our predictivity criterion, 11 of these 15
textures are predictive with respect to one or more of the
three light quark masses $m_u$, $m_d$ and $m_s$; the masses of
the heavy quarks and the Wolfenstein parameters cannot be predicted.

At first sight, the last sentence seems
to exclude approximate relations of the form of
equation~(\ref{c}) or, more generally, relations of the
form\footnote{For the sake of simplicity, in this paragraph we confine
  ourselves to three observables.}
\begin{equation}\label{relation}
f\left(
\frac{m_1}{m_2},\, W
\right) = 0,
\end{equation}
where $f$ denotes a function specified by the particular form of the texture,
$m_1$ and $m_2$ are quark masses and $W$ is one of the Wolfenstein parameters.
Clearly, if a relation of the type of equation~(\ref{relation})
follows from a texture, then $W$ is predicted by $m_1/m_2$, but
mathematically we can turn this conclusion around and say that 
$m_1$ is predicted by $m_2$ and $W$ or $m_2$ is predicted by $m_1$ and $W$.
However, \emph{numerically} this will in general not be the case
because of the different relative errors 
$\sigma_j/\hspace{1pt}\overline{\mathcal{O}}_j$ of the observables. 
From table~\ref{input} one finds that the relative errors
of the light quark masses 
$m_u$, $m_d$ and $m_s$
are between $5\%$ and $40\%$, the errors of the heavy quark
masses are between $1\%$ and $3\%$ and the errors of the Wolfenstein parameters
range from $0.3\%$ for $\lambda$ to $15\%$ for $\bar{\rho}$. 
For instance, 
for the famous relation~(\ref{c}),
varying $m_d$ and $m_s$ around their experimental mean values within
ranges of the order of magnitude of their respective experimental errors, 
one may well obtain values for $\lambda \equiv \sin\theta_c$ which, due to
the small relative error
$\sigma_\lambda/\lambda \sim 0.3\%$, may lie more than 10 sigmas off its
mean value, \textit{i.e.}\ $\Delta(\lambda)>100$.
Conversely, fixing $\lambda$ and one of the masses gives a prediction
for the 
second
mass which, since the relative errors of $m_d$ and $m_s$
are much larger, 
may very well be within ten sigmas of the experimental
value, \textit{i.e.} $\Delta(m_q)<100$ for $q=d,s$.
To summarize, even if there is a relation of the form~(\ref{relation}),
the predictivity analysis will in general not 
detect all involved observables.
This has to be kept in mind in the assessment of the results displayed
in table~\ref{symmetric}.

All of the 15 textures for symmetric quark mass matrices
have four $(S_1,\ldots,S_7,S_{11},S_{14})$ or five
$(S_8,S_9,S_{10},S_{12},S_{13},S_{15})$ independent texture zeros,
the most predictive one 
being\footnote{Note that for the mass matrices of equation~(\ref{S10})
  we find numerically 
  $| (M_d)_{22} | \simeq m_b$, 
  $| (M_u)_{22} | \simeq m_t$. For achieving the usual ordering with
  large elements having higher indices, one
  has to apply a permutation $2 \leftrightarrow 3$ to the indices.}
\begin{equation}\label{S10}
S_{10}:\quad
M_d\sim
\begin{pmatrix}
0 & 0 & 1 \\
0 & 2 & 1 \\
1 & 1 & 1
\end{pmatrix},
\quad
M_u\sim 
\begin{pmatrix}
0 & 1 & 0 \\
1 & 2 & 0 \\
0 & 0 & 1
\end{pmatrix},
\end{equation}
which shows predictive power with respect to all of the
three light quark masses.
In the light of the discussion above, this does not mean that the
texture $S_{10}$ has three \emph{independent} predictions. Indeed, in
this case there are only two independent ones, which can be formulated as
$\sin\theta_{12} \simeq \sqrt{m_d/m_s}$, \textit{i.e.}\ equation~(\ref{c}), and
$|U_{ub}| \equiv \sin\theta_{13} \simeq \sqrt{m_u/m_t}$.

\renewcommand{\arraystretch}{1.1}
\begin{table}
\begin{footnotesize}
\begin{center}
\input{classes_symmetric.tex}
\end{center}
\end{footnotesize}
\caption{The 15 maximally restrictive textures in symmetric quark mass matrices.
A matrix entry $0$ denotes a texture zero, and entries $1$ and $2$ stand for
real positive and complex parameters, respectively. Several of these textures
are predictive with respect to some of the light quark masses.
\label{symmetric}}
\end{table}

We emphasize once more that, after the consideration of general quark
mass matrices, we have investigated \emph{symmetric} (but not
\emph{Hermitian}) mass matrices, in which context we have discovered
six viable textures with five texture zeros---see
table~\ref{symmetric}. It is interesting to compare these six
textures with the five viable \emph{Hermitian} textures with five
texture zeros discussed in the literature~\cite{rrr,ponce}. For this
comparison we use the table in~\cite{ponce} where the five Hermitian
patterns I-V are listed and check if there are corresponding patterns
in our table~\ref{symmetric}, with zeros in corresponding places after
suitable weak-basis permutations. We find the correspondences 
$\mathrm{II} \sim S_9$,
$\mathrm{III} \sim S_8$,
$\mathrm{IV} \sim S_{15}$ and
$\mathrm{V} \sim S_{10}$,
while pattern~I 
has no correspondence to
viable texture zeros in symmetric
quark mass matrices. This comparison reveals the fundamental difference
between texture zeros in symmetric and Hermitian
mass matrices.\footnote{Note that in~\cite{ponce} the
Hermitian mass matrices I-V are studied at $\mu=M_Z$
just as in this paper. Hence, 
the results can directly be compared.}

\paragraph{Summary:} 
In this 
paper we have performed a systematic and complete analysis of
texture zeros in general and symmetric quark mass matrices. 
Among all the possible texture zeros in
general quark mass matrices, we identified the 27 maximally restrictive
classes---see table~\ref{general}---which, however, do not show
predictive power with respect to any of the quark masses and mixing parameters. 
This is very similar to the situation of Dirac neutrinos, where texture zeros
are predictive 
at most with respect to the smallest neutrino mass
and, in one case, also to the Dirac phase $\delta$ of the lepton
mixing matrix~\cite{ludl2014}. 
In other words, pure Abelian flavour symmetries effective at the
electroweak scale, \textit{i.e.}\ texture zeros but no further
restrictions on the quark mass matrices, do not seem to contribute to
the solution of the mass and mixing problem in the quark sector.

The case of texture zeros in symmetric quark mass matrices is more promising,
since there the majority of the 15 maximally restrictive textures has 
some predictive power---see table~\ref{symmetric}.
This may be compared to the case of Majorana neutrinos,
where the neutrino mass matrix is symmetric. There the maximally restrictive
textures are also more predictive than for the Dirac neutrino
case~\cite{ludl2014}.

\paragraph{Acknowledgments:} This work is supported by the Austrian
Science Fund (FWF), Project No.\ P~24161-N16. 
P.O.L.\ thanks 
Helmut Moser for his tireless servicing of our research group's computer cluster
on which the computations for this work were performed.

\end{document}

%% file: classes_general.tex
\begin{tabular}{|lll|lll|lll|}
\hline
& \multicolumn{1}{c}{$M_d$} & \multicolumn{1}{c|}{$M_u$} &
& \multicolumn{1}{c}{$M_d$} & \multicolumn{1}{c|}{$M_u$} &
& \multicolumn{1}{c}{$M_d$} & \multicolumn{1}{c|}{$M_u$} \\
\hline
$G_{1}$
 &
$\begin{pmatrix}
0 & 1 & 1 \\
0 & 1 & 2 \\
1 & 0 & 1
\end{pmatrix}$
 &
$\begin{pmatrix}
0 & 0 & 1 \\
0 & 1 & 0 \\
1 & 0 & 0
\end{pmatrix}$
 &
$G_{2}$
 &
$\begin{pmatrix}
0 & 0 & 1 \\
0 & 1 & 0 \\
1 & 1 & 2
\end{pmatrix}$
 &
$\begin{pmatrix}
0 & 0 & 1 \\
0 & 1 & 0 \\
1 & 0 & 1
\end{pmatrix}$
 &
$G_{3}$
 &
$\begin{pmatrix}
0 & 0 & 1 \\
0 & 1 & 0 \\
1 & 1 & 2
\end{pmatrix}$
 &
$\begin{pmatrix}
0 & 0 & 1 \\
0 & 1 & 1 \\
1 & 0 & 0
\end{pmatrix}$
\\
$G_{4}$
 &
$\begin{pmatrix}
0 & 0 & 1 \\
0 & 1 & 0 \\
1 & 2 & 1
\end{pmatrix}$
 &
$\begin{pmatrix}
0 & 0 & 1 \\
1 & 1 & 0 \\
0 & 1 & 0
\end{pmatrix}$
 &
$G_{5}$
 &
$\begin{pmatrix}
0 & 0 & 1 \\
0 & 1 & 1 \\
1 & 0 & 2
\end{pmatrix}$
 &
$\begin{pmatrix}
0 & 0 & 1 \\
0 & 1 & 0 \\
1 & 0 & 1
\end{pmatrix}$
 &
$G_{6}$
 &
$\begin{pmatrix}
0 & 0 & 1 \\
0 & 1 & 1 \\
1 & 0 & 2
\end{pmatrix}$
 &
$\begin{pmatrix}
0 & 0 & 1 \\
0 & 1 & 0 \\
1 & 1 & 0
\end{pmatrix}$
\\
$G_{7}$
 &
$\begin{pmatrix}
0 & 0 & 1 \\
0 & 1 & 2 \\
1 & 0 & 1
\end{pmatrix}$
 &
$\begin{pmatrix}
0 & 1 & 1 \\
0 & 0 & 1 \\
1 & 0 & 0
\end{pmatrix}$
 &
$G_{8}$
 &
$\begin{pmatrix}
0 & 0 & 1 \\
0 & 1 & 1 \\
1 & 2 & 0
\end{pmatrix}$
 &
$\begin{pmatrix}
0 & 0 & 1 \\
0 & 1 & 0 \\
1 & 0 & 1
\end{pmatrix}$
 &
$G_{9}$
 &
$\begin{pmatrix}
0 & 0 & 1 \\
0 & 1 & 1 \\
1 & 2 & 0
\end{pmatrix}$
 &
$\begin{pmatrix}
0 & 0 & 1 \\
0 & 1 & 0 \\
1 & 1 & 0
\end{pmatrix}$
\\
$G_{10}$
 &
$\begin{pmatrix}
0 & 0 & 1 \\
0 & 1 & 2 \\
1 & 1 & 0
\end{pmatrix}$
 &
$\begin{pmatrix}
0 & 0 & 1 \\
0 & 1 & 1 \\
1 & 0 & 0
\end{pmatrix}$
 &
$G_{11}$
 &
$\begin{pmatrix}
0 & 0 & 1 \\
0 & 1 & 1 \\
1 & 2 & 0
\end{pmatrix}$
 &
$\begin{pmatrix}
0 & 0 & 1 \\
1 & 1 & 0 \\
0 & 1 & 0
\end{pmatrix}$
 &
$G_{12}$
 &
$\begin{pmatrix}
0 & 0 & 1 \\
0 & 1 & 2 \\
1 & 1 & 0
\end{pmatrix}$
 &
$\begin{pmatrix}
0 & 1 & 1 \\
0 & 0 & 1 \\
1 & 0 & 0
\end{pmatrix}$
\\
$G_{13}$
 &
$\begin{pmatrix}
0 & 0 & 1 \\
0 & 1 & 1 \\
1 & 2 & 0
\end{pmatrix}$
 &
$\begin{pmatrix}
0 & 1 & 1 \\
1 & 0 & 0 \\
0 & 0 & 1
\end{pmatrix}$
 &
$G_{14}$
 &
$\begin{pmatrix}
0 & 0 & 1 \\
0 & 1 & 0 \\
1 & 0 & 2
\end{pmatrix}$
 &
$\begin{pmatrix}
0 & 0 & 1 \\
0 & 1 & 0 \\
1 & 1 & 1
\end{pmatrix}$
 &
$G_{15}$
 &
$\begin{pmatrix}
0 & 0 & 1 \\
0 & 1 & 0 \\
1 & 0 & 2
\end{pmatrix}$
 &
$\begin{pmatrix}
0 & 0 & 1 \\
0 & 1 & 1 \\
1 & 0 & 1
\end{pmatrix}$
\\
$G_{16}$
 &
$\begin{pmatrix}
0 & 0 & 1 \\
0 & 1 & 0 \\
1 & 0 & 2
\end{pmatrix}$
 &
$\begin{pmatrix}
0 & 0 & 1 \\
0 & 1 & 1 \\
1 & 1 & 0
\end{pmatrix}$
 &
$G_{17}$
 &
$\begin{pmatrix}
0 & 0 & 1 \\
0 & 1 & 0 \\
1 & 0 & 2
\end{pmatrix}$
 &
$\begin{pmatrix}
0 & 0 & 1 \\
1 & 1 & 0 \\
0 & 1 & 1
\end{pmatrix}$
 &
$G_{18}$
 &
$\begin{pmatrix}
0 & 0 & 1 \\
0 & 1 & 0 \\
1 & 0 & 2
\end{pmatrix}$
 &
$\begin{pmatrix}
0 & 0 & 1 \\
1 & 1 & 1 \\
0 & 1 & 0
\end{pmatrix}$
\\
$G_{19}$
 &
$\begin{pmatrix}
0 & 0 & 1 \\
0 & 1 & 0 \\
1 & 0 & 2
\end{pmatrix}$
 &
$\begin{pmatrix}
0 & 1 & 1 \\
0 & 0 & 1 \\
1 & 0 & 1
\end{pmatrix}$
 &
$G_{20}$
 &
$\begin{pmatrix}
0 & 0 & 1 \\
0 & 1 & 0 \\
1 & 0 & 2
\end{pmatrix}$
 &
$\begin{pmatrix}
0 & 1 & 1 \\
0 & 0 & 1 \\
1 & 1 & 0
\end{pmatrix}$
 &
$G_{21}$
 &
$\begin{pmatrix}
0 & 0 & 1 \\
0 & 1 & 0 \\
1 & 0 & 2
\end{pmatrix}$
 &
$\begin{pmatrix}
0 & 1 & 1 \\
1 & 0 & 0 \\
1 & 0 & 1
\end{pmatrix}$
\\
$G_{22}$
 &
$\begin{pmatrix}
0 & 0 & 1 \\
0 & 1 & 0 \\
1 & 0 & 2
\end{pmatrix}$
 &
$\begin{pmatrix}
0 & 1 & 1 \\
1 & 0 & 1 \\
0 & 0 & 1
\end{pmatrix}$
 &
$G_{23}$
 &
$\begin{pmatrix}
0 & 0 & 1 \\
0 & 1 & 0 \\
1 & 0 & 2
\end{pmatrix}$
 &
$\begin{pmatrix}
0 & 1 & 1 \\
1 & 0 & 1 \\
0 & 1 & 0
\end{pmatrix}$
 &
$G_{24}$
 &
$\begin{pmatrix}
0 & 0 & 1 \\
0 & 1 & 0 \\
1 & 0 & 2
\end{pmatrix}$
 &
$\begin{pmatrix}
0 & 1 & 1 \\
1 & 0 & 1 \\
1 & 0 & 0
\end{pmatrix}$
\\
$G_{25}$
 &
$\begin{pmatrix}
0 & 0 & 1 \\
0 & 1 & 0 \\
1 & 0 & 2
\end{pmatrix}$
 &
$\begin{pmatrix}
1 & 1 & 1 \\
0 & 0 & 1 \\
0 & 1 & 0
\end{pmatrix}$
 &
$G_{26}$
 &
$\begin{pmatrix}
0 & 0 & 1 \\
0 & 1 & 0 \\
1 & 0 & 0
\end{pmatrix}$
 &
$\begin{pmatrix}
0 & 0 & 1 \\
1 & 1 & 1 \\
0 & 1 & 2
\end{pmatrix}$
 &
$G_{27}$
 &
$\begin{pmatrix}
0 & 0 & 1 \\
0 & 1 & 0 \\
1 & 0 & 0
\end{pmatrix}$
 &
$\begin{pmatrix}
0 & 1 & 1 \\
0 & 1 & 2 \\
1 & 0 & 1
\end{pmatrix}$
\\
\hline
\end{tabular}

%% file: classes_symmetric.tex
\begin{tabular}{|llll|llll|}
\hline
& \multicolumn{1}{c}{$M_d$} & \multicolumn{1}{c}{$M_u$} & \multicolumn{1}{c|}{predicted obs.} &
& \multicolumn{1}{c}{$M_d$} & \multicolumn{1}{c}{$M_u$} & \multicolumn{1}{c|}{predicted obs.} \\
\hline
$S_{1}$
 &
$\begin{pmatrix}
0 & 1 & 1 \\
1 & 1 & 2 \\
1 & 2 & 2
\end{pmatrix}$
 &
$\begin{pmatrix}
1 & 0 & 0 \\
0 & 0 & 1 \\
0 & 1 & 2
\end{pmatrix}$
 &
$m_d$
 &
$S_{2}$
 &
$\begin{pmatrix}
0 & 1 & 1 \\
1 & 1 & 2 \\
1 & 2 & 2
\end{pmatrix}$
 &
$\begin{pmatrix}
1 & 0 & 0 \\
0 & 1 & 0 \\
0 & 0 & 1
\end{pmatrix}$
 &
$m_d$
\\
$S_{3}$
 &
$\begin{pmatrix}
0 & 1 & 1 \\
1 & 0 & 1 \\
1 & 1 & 2
\end{pmatrix}$
 &
$\begin{pmatrix}
1 & 0 & 0 \\
0 & 1 & 2 \\
0 & 2 & 2
\end{pmatrix}$
 &
$m_d$
 &
$S_{4}$
 &
$\begin{pmatrix}
0 & 0 & 1 \\
0 & 2 & 2 \\
1 & 2 & 2
\end{pmatrix}$
 &
$\begin{pmatrix}
0 & 1 & 1 \\
1 & 0 & 1 \\
1 & 1 & 2
\end{pmatrix}$
 &
---
\\
$S_{5}$
 &
$\begin{pmatrix}
0 & 0 & 1 \\
0 & 2 & 2 \\
1 & 2 & 2
\end{pmatrix}$
 &
$\begin{pmatrix}
1 & 0 & 1 \\
0 & 0 & 1 \\
1 & 1 & 2
\end{pmatrix}$
 &
---
 &
$S_{6}$
 &
$\begin{pmatrix}
0 & 0 & 1 \\
0 & 2 & 2 \\
1 & 2 & 2
\end{pmatrix}$
 &
$\begin{pmatrix}
1 & 1 & 0 \\
1 & 2 & 1 \\
0 & 1 & 0
\end{pmatrix}$
 &
---
\\
$S_{7}$
 &
$\begin{pmatrix}
0 & 0 & 1 \\
0 & 1 & 1 \\
1 & 1 & 2
\end{pmatrix}$
 &
$\begin{pmatrix}
1 & 0 & 0 \\
0 & 1 & 2 \\
0 & 2 & 2
\end{pmatrix}$
 &
$m_d$
 &
$S_{8}$
 &
$\begin{pmatrix}
0 & 0 & 1 \\
0 & 2 & 2 \\
1 & 2 & 2
\end{pmatrix}$
 &
$\begin{pmatrix}
0 & 0 & 1 \\
0 & 1 & 1 \\
1 & 1 & 0
\end{pmatrix}$
 &
$m_d$
\\
$S_{9}$
 &
$\begin{pmatrix}
0 & 0 & 1 \\
0 & 1 & 1 \\
1 & 1 & 2
\end{pmatrix}$
 &
$\begin{pmatrix}
0 & 0 & 1 \\
0 & 1 & 0 \\
1 & 0 & 2
\end{pmatrix}$
 &
$m_d$
 &
$S_{10}$
 &
$\begin{pmatrix}
0 & 0 & 1 \\
0 & 1 & 1 \\
1 & 1 & 2
\end{pmatrix}$
 &
$\begin{pmatrix}
0 & 1 & 0 \\
1 & 2 & 0 \\
0 & 0 & 1
\end{pmatrix}$
 &
$m_u$, $m_d$, $m_s$
\\
$S_{11}$
 &
$\begin{pmatrix}
0 & 1 & 1 \\
1 & 1 & 0 \\
1 & 0 & 2
\end{pmatrix}$
 &
$\begin{pmatrix}
1 & 0 & 0 \\
0 & 1 & 2 \\
0 & 2 & 2
\end{pmatrix}$
 &
$m_d$
 &
$S_{12}$
 &
$\begin{pmatrix}
0 & 0 & 1 \\
0 & 2 & 2 \\
1 & 2 & 0
\end{pmatrix}$
 &
$\begin{pmatrix}
0 & 0 & 1 \\
0 & 1 & 1 \\
1 & 1 & 2
\end{pmatrix}$
 &
$m_d$
\\
$S_{13}$
 &
$\begin{pmatrix}
0 & 0 & 1 \\
0 & 2 & 2 \\
1 & 2 & 0
\end{pmatrix}$
 &
$\begin{pmatrix}
0 & 1 & 0 \\
1 & 1 & 1 \\
0 & 1 & 2
\end{pmatrix}$
 &
$m_d$, $m_s$
 &
$S_{14}$
 &
$\begin{pmatrix}
0 & 0 & 1 \\
0 & 1 & 0 \\
1 & 0 & 2
\end{pmatrix}$
 &
$\begin{pmatrix}
1 & 0 & 1 \\
0 & 1 & 2 \\
1 & 2 & 2
\end{pmatrix}$
 &
---
\\
$S_{15}$
 &
$\begin{pmatrix}
0 & 0 & 1 \\
0 & 1 & 0 \\
1 & 0 & 2
\end{pmatrix}$
 &
$\begin{pmatrix}
0 & 1 & 0 \\
1 & 1 & 1 \\
0 & 1 & 2
\end{pmatrix}$
 &
$m_d$, $m_s$
&
&
&
&
\\
\hline
\end{tabular}